\documentclass[letterpaper,12pt]{JHEP3}

\usepackage{amsmath}
\usepackage{amssymb}
\usepackage{graphicx}

\def\Re{{\cal R \mskip-4mu \lower.1ex \hbox{\it e}\,}}
\def\Im{{\cal I \mskip-5mu \lower.1ex \hbox{\it m}\,}}
\def\ie{{\it i.e.}}

\def\tev{\,{\rm TeV}}
\def\gev{\,{\rm GeV}}
\def\mev{\,{\rm MeV}}
\def\to{\rightarrow}

\newskip\zatskip \zatskip=0pt plus0pt minus0pt
\def\matth{\mathsurround=0pt}
\def\lsim{\mathrel{\mathpalette\atversim<}}

\def\atversim#1#2{\lower0.7ex\vbox{\baselineskip\zatskip\lineskip\zatskip
  \lineskiplimit 0pt\ialign{$\matth#1\hfil##\hfil$\crcr#2\crcr\sim\crcr}}}

\author{Ben Lillie\\Stanford Linear Accelerator Center\\Stanford CA 94309,
USA\\ \email{lillieb@slac.stanford.edu}\thanks{Work supported by the Department of Energy, Contract DE-AC03-76SF00515}}
\title{Yukawa Hierarchies From Extra Dimensions With Small FCNC}
\preprint{SLAC-PUB-10094}

\abstract{We investigate a class of extra dimensional models where all of the
Standard Model fermions are localized to a single fixed point in an 
$S_1/Z_2$ orbifold, and each species is localized with an exponential
wavefunction with a different
width. We show
that this naturally generates Yukawa hierarchies of the size present in the
Standard Model, and we find a set of model parameters that reproduces the
observed masses and mixings to experimental accuracy. In addition, the
dominant
constraints, arising from flavor changing neutral currents, are shown to restrict the
compactification scale to be $1/R \ge 2-5 \tev$, which is a much less stringent
constraint than in similar extra dimensional models of the Yukawa hierarchy.}

\begin{document}

\section{Introduction}

Recently there has been much interest in the possibility that there exist
extra dimensions much larger than the Planck
scale \cite{Antoniadis:1990ew,Lykken:1996fj,Witten:1996mz,Horava:1996ma,Horava:1996qa,Caceres:1997is}.
There are many reasons for this excitement, particularly the fact that
having large dimensions accessible to gravity may allow a 
solution of the hierarchy problem associated with the Higgs
mass \cite{Arkani-Hamed:1998rs,Antoniadis:1998ig,Arkani-Hamed:1998nn,Randall:1999ee,Randall:1999vf}.
However, extra dimensions may also be able to address other puzzles left
unexplained in the Standard Model, such as why there are three families
\cite{Dobrescu:2001ae}, or the origin of dark matter
\cite{Appelquist:2000nn,Servant:2002aq,Servant:2002hb,Cheng:2002ej}.

In particular, it was noticed by Arkani-Hamed and Schmaltz (AS) that
localizing the standard model fermion fields in an extra dimension
could solve the other serious hierarchy problem, that of the relative size
of the fermion masses \cite{Arkani-Hamed:1999dc}. This is accomplished by
assigning a universal Yukawa coupling in the higher dimensional theory,
but separating the left and right handed components of the effective 4D
fermions in the additional dimension. This is accomplished by
localizing the zero modes to Gaussian profiles, and separating the centers
of the Gaussians. The 4D Yukawa couplings, and hence fermion masses, are
proportional to integrals of products of these zero modes over the compact
dimension. They are thus
suppressed by exponentially small wavefunction overlaps. Realizations of
this model that match the Standard Model masses and mixing parameters
have been produced \cite{Grossman:2002pb,Mirabelli:1999ks,Chang:2002ww,
DelAguila:2001pu,Kaplan:2001ga,Branco:2000rb}.
There also exist possibilities for observing the effects of this
localization at future
colliders, such as the production of Kaluza-Klein
excitations of the Standard Model particles.
Current results from the direct production of KK states. and
precision electro-weak measurements restrict the size of
dimensions accessible to Standard Model fields to be $R \lsim {\rm few}
\tev^{-1}$ \cite{Rizzo:1999br,Masip:1999mk,Marciano:1999ih,Hewett:2002hv}.
The separation between fields also allows for detection of some novel effects
\cite{Arkani-Hamed:1999za,Rizzo:2001cy}.

However, split fermion models also generate tree-level
flavor-changing neutral currents (FCNC). Strict bounds from these are hard
to obtain due to the large number of model parameters, but in general one
finds $R \lsim 400
\tev^{-1}$ \cite{Chang:2002ww,Delgado:1999sv,Abel:2003fk,Lillie:2003yz}.
Without miraculous cancellations, this can be evaded only by making very
small the ratio $\rho = \sigma/R$, where $\sigma$ is the
width of the localized fermions, thus introducing a new hierarchy.

In this paper we present a different model based on similar
ideas. Instead of fixed width fermions localized at different points, we
consider variable width fermions localized at the same point. We will show
that this model can produce the observed fermion mass hierarchy and mixing
angles, including the CP violating phase while generating much smaller
FCNC. This model has the additional virtue suppressing any proton decay
operators. The resulting constraints on the 
compactification scale turn out to be a factor of 30
smaller than similar constraints on the AS model, and can be as low as
$1/R \ge 2 \tev$. This is low enough to raise the exciting possibility that they could be
within reach of the next generation of colliders, and that the model could
be embedded in a more encompassing one that also solves the Higgs mass
hierarchy problem. This is similar to the
proposal of Kaplan and Tait \cite{Kaplan:2001ga}, although they still
required the left and right handed fermions to be separated from each
other. This idea is also similar to a model studied in the case of a
warped extra dimension \cite{Huber:2003tu}.

This paper is organized as follows. In section \ref{section:yukawa} we show
how variable widths can explain the Yukawa hierarchy. A specific set of
model parameters that reproduce the observed fermion masses and mixing
angles is shown in section \ref{section:standardmodel}. In section
\ref{section:fcnc} we consider the constraints from FCNC. Section
\ref{section:mechanisms} discusses possible mechanisms for implementing
the model, and section \ref{section:conclusion} concludes.

\section{Yukawa Hierarchies}\label{section:yukawa}

We propose the configuration where different fermion
flavors are localized in the extra-dimensional bulk as exponentially
localized wavefunctions with different widths,
as illustrated in 
Fig. \ref{fig:cartoon}. For most of this paper we consider this to be a
phenomenological ans\"atz. A discussion of what types of models may
produce this picture is given in section \ref{section:mechanisms}.
This paper will focus on models with a single compact extra dimension,
which we take to be the interval $[0,R]$, with the edges of the
additional dimension being defined either by a brane configuration or by
an orbifolding condition. 

We begin with a general configuration that allows each fermion species to
be localized with different widths to a specific point.
A fermion is taken to be localized at $x_0$ with width $\sigma$ if it
has a profile $\psi 
\propto e^{-|x-x_0|^{a}/\sigma^a}$. The power $a$ is determined by the
localization mechanism, and represents a phenomenological parameterization
of the unknown mechanism. In the AS model,
for instance, fermions are localized with Gaussian profiles, so $a=2$; in
the Randall-Sundrum (RS) scenario, or in brane-worlds with bulk mass
terms, fermions are
localized to exponentials, so $a=1$. For simplicity, we take the
localization point to be $x_0 = 0$.
The normalized wavefunction for the $i$-th fermion is
\begin{gather}
\psi_i = \frac{2^{1/2a}}{\sqrt{\sigma_i}} \frac{1}{\sqrt{\alpha_a(R,\sigma)}}
e^{-y^a/\sigma_i^a},\label{eq:profile}
\end{gather}
where we have defined
\begin{gather}
\alpha_a(R,\sigma) = \int_0^{\frac{R}{\sigma}} dz\ e^{-z^a}.
\end{gather}
This integral will cancel to good approximation in the following
computations of the coupling constants.

\FIGURE[t]{
\centerline{
\includegraphics[width=12cm,height=5cm,angle=0]{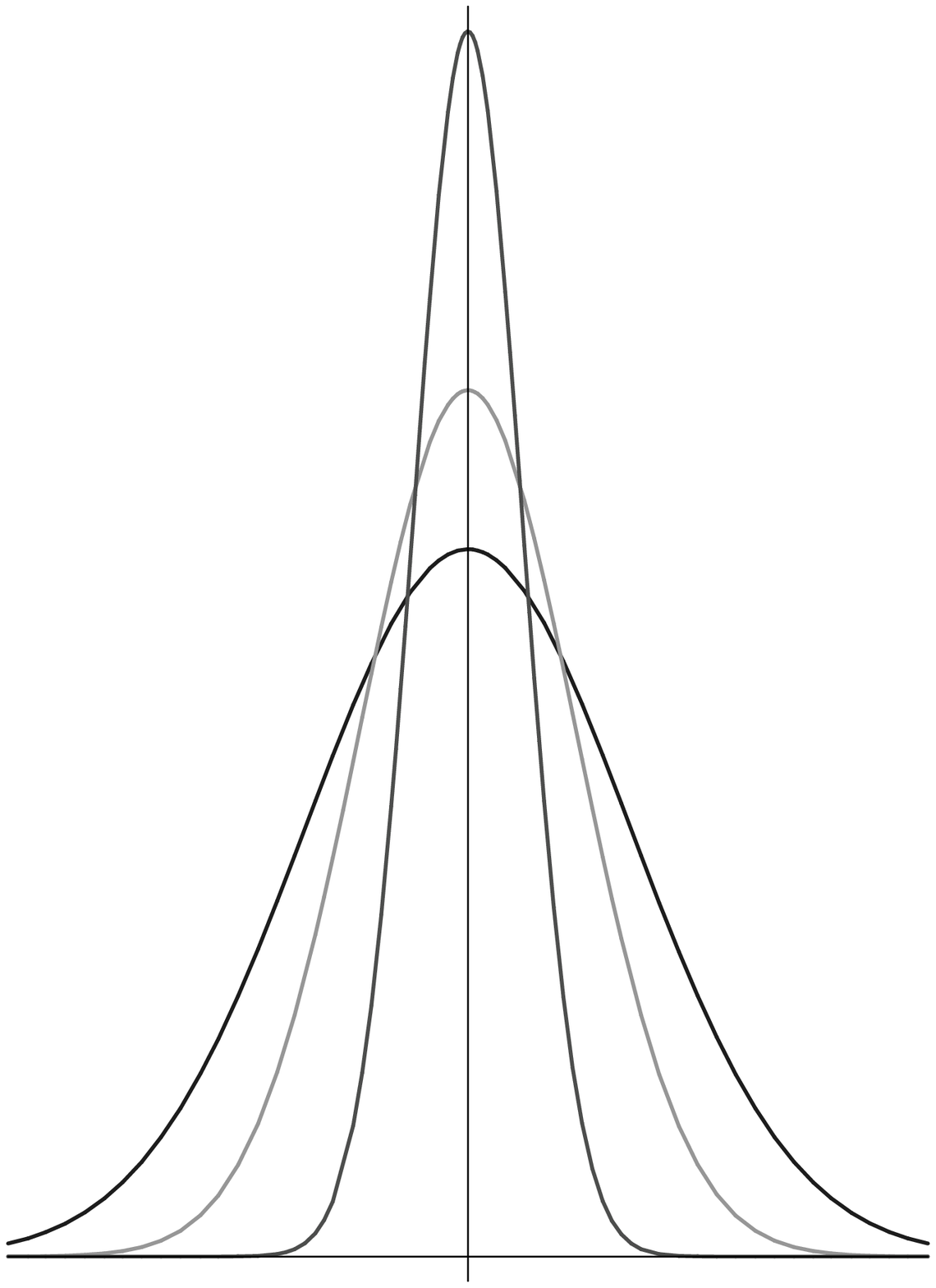}}
\caption{\small
Illustration of the idea of variable width fermions localized to a
single point.
}
\label{fig:cartoon}}

The gauge boson wavefunctions will also generally depend on the model
details. For the moment the only important point is that the Higgs zero
mode be flat in the extra dimension, so $h^{(0)}(y) = 1/\sqrt{R}$, where
$h$ is the Higgs field and $h^{(0)}$ is the zero mode of the Kaluza-Klein
expansion. This zero mode corresponds to the Standard Model Higgs, the vev
of which is responsible for electro-weak symmetry breaking and generating 4D
fermion masses.
The 5D Yukawa couplings for the fermions carry mass dimension $-1/2$. In the
dimensionally reduced theory this scale is given by the compactification
scale $R$. We therefore write the 5D Yukawa terms as
\begin{gather}
{\cal L}^{5D}_{\rm Yukawa} = \lambda_{5,ij} \sqrt{R}\, h \bar\psi_i\psi_j.
\end{gather}
The dimensionally reduced 4D Lagrangian is obtained by integrating over the
compact dimension.
The 4D Yukawa couplings are then given, in the physically appropriate
approximation that $R$ is much larger than any of the widths, by
\begin{align}
\lambda_{4,ij} &= 
\lambda_{5,ij} \sqrt{R}\int_{0}^{R}dy\ h^{(0)}(y)\ \bar\psi_i(y)\ \psi_j(y)\notag\\
& = \lambda_{5,ij} \sqrt{R}\left(\frac{2^{1/a}\sqrt{\sigma_i \sigma_j}}{(\sigma_i^a +
\sigma_j^a)^{1/a}}\right)\label{eq:yukawacouplings}.
\end{align}
This can be rewritten as
\begin{gather}
\lambda_{4,ij} = \lambda_{5,ij}
\sqrt{R}
\left(2^{1/a}
\left(
\left(
\frac{\sigma_i}{\sigma_j}
\right)^{a/2}
+ \left(
\frac{\sigma_j}{\sigma_i}
\right)^{a/2}
\right)^{-1/a}
\right).
\end{gather}
The factor in parenthesis is determined by the wavefunction
overlap. It is unity when $\sigma_i = \sigma_j$,
and does not change
dramatically as the widths vary. If we take all the 5D Yukawa couplings
to be the same, $\lambda_{5,ij} = \lambda_5$ we expect this model to have 4D
Yukawa matrix elements given by $\lambda_{4,ij} \approx \lambda_5$,
and essentially equal with small differences due to the different widths. This is a realization
of the ``democratic'' scenario of fermion mass
generation \cite{Fritzsch:1999ee}. This scenario relies on the 
observation of the singular values of the matrix
\begin{gather}
A = \begin{pmatrix}
1 & 1 & 1 \\
1 & 1 & 1 \\
1 & 1 & 1
    \end{pmatrix}
\xrightarrow{{\rm diag}}
\begin{pmatrix}
3 & 0 & 0 \\
0 & 0 & 0 \\
0 & 0 & 0
    \end{pmatrix}.
\end{gather}
That is, there is one large singular value and two vanishing ones. If one then
perturbs the values of the elements in $A$, the two zero diagonal values become finite, but
small. When $A$ is a Yukawa matrix this gives a natural hierarchy in the
resulting masses. This is in stark contrast to the AS scenario, where the
hierarchy is generated by small elements directly in the Yukawa matrices.

We have studied the expected size of the resulting Yukawa hierarchy by
performing random trials. We drew six widths for the 5D fermion wavefunctions from
the interval $[1,3]$ and computed the physical mass spectrum for the case
$a=2$. The interval was chosen so that all widths are of the same order of
magnitude; the overall scale is irrelevant. The size of the
hierarchy is taken to be $\log_{10}(m_1/m_3)$, where $m_1$ is the largest mass,
and $m_3$ the smallest. Fig. \ref{fig:hierarchy}  shows a histogram of the
resulting hierarchies for $10^{5}$ random trials. We see that hierarchies
of size $3$ to $5$ are generic.

\FIGURE[t]{\centerline{
\includegraphics[width=12cm]{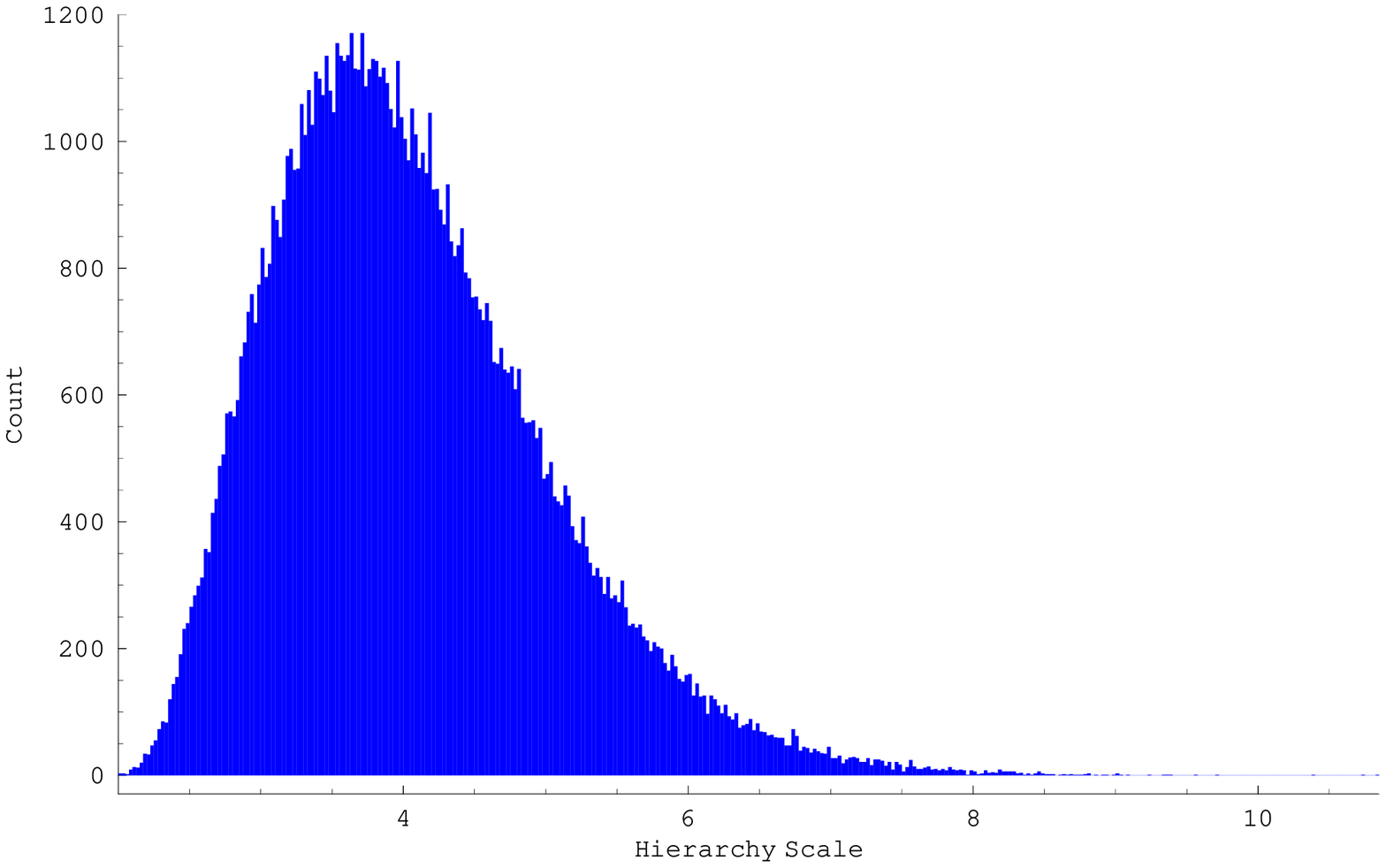}
}
\caption{\small
Histogram of the expected size of the Yukawa hierarchy for $10^{5}$ random
configurations of fermion widths, with the widths drawn from the interval
$[1,3]$. The hierarchy scale is defined as $\log_{10}(m_1/m_3)$, where
$m_1$($m_3$) is the largest (smallest) mass resulting from that
configuration.
}
\label{fig:hierarchy}}

It is important to note that Fig \ref{fig:hierarchy} shows the hierarchy
between the masses resulting from a single Yukawa matrix; \ie, 
between fermions with the same quantum numbers. In this democratic scheme
the largest singular value (and hence mass) is always of the same order of
magnitude. Therefore it is impossible to
generate the the large ratios $m_t/m_b$ or $m_t/m_\tau$ simply by varying
the widths. We speculate on how to obtain these ratios within the model in
section \ref{section:mechanisms}; for now we simply take the five
dimensional Yukawa couplings $\lambda_5^u,\,\lambda_5^d,\,\lambda_5^e$ to be
different.

Finally, note that while Fig. \ref{fig:hierarchy} was computed for $a=2$,
similar results hold for any value of $a$. The important point is that the mass
matrix is almost universal, and that small perturbations are generated by
slightly different zero mode profiles. In addition, the Higgs profile need not
be perfectly flat. As long as the variation in the Higgs 5D wavefunction
is slow across the width of the fermions the same pattern will emerge.

\section{Standard Model parameters}
\label{section:standardmodel}

For illustration, we will now obtain a set of model parameters that generates
the observed fermion masses and mixings of the Standard Model to within
experimental accuracy. For now, our model contains the following: There is
the localization mechanism parameter $a$, which will be fixed to 2 in this
section. There
are the three Yukawa couplings
$\lambda_5^u,\,\lambda_5^d,\,\lambda_5^e$. For the quark sector, there are
nine widths, one each for the three left handed quark doublets, $Q_i$, the
three right handed up singlets, $\bar{u}_i$, and the three right handed down
singlets,
$\bar{d}_i$. Only eight of the 
widths are independent because the Yukawa matrix elements from Eq.
\ref{eq:yukawacouplings} only depend on the ratios of widths. For the
lepton sector there are at least another six widths, for the doublets
$L_i$, and the charged singlets $e^+_i$.
There are possibly three more
depending on whether right-handed neutrinos are included. For the purposes
of this paper we ignore the neutrinos and only match the charged lepton
masses to their observed values. Finally, it is necessary to include two arbitrary, but small,
phases in one of the quark Yukawa matrices to be able to match all the CKM
mixing angles and incorporate CP violation.

As a proof of principle, we have performed a search of the parameter space
and located a configuration that reproduces the Standard Model
parameters. For this
search we used the values of quark masses and mixing angles from
\cite{Hagiwara:2002fs}. The fermion masses were evaluated at the common scale
$m_t$ using factors given in \cite{Mirabelli:1999ks}. The values of the
three five dimensional Yukawa couplings were obtained by matching to the large
third generation masses (so $\lambda_5^u = \sqrt{2}m_t/v$, etc. where $v$
is the Higgs vev). The smaller fermion masses are then obtained by
diagonalizing the Yukawa matrices generated from
Eq. (\ref{eq:yukawacouplings}). The mixing angles can then be obtained
from the relation
\begin{gather}
V_{\rm ckm} = V_L^{(u)\dagger}V_L^{(d)},
\end{gather}
where the $V_L^{(u,d)}$ are the unitary matrices that rotate the
left-handed $(u,d)$ quark fields to diagonalize the Yukawa matrices. The
real part of the CKM matrix was parameterized by matching to the magnitude
of the three entries above the diagonal, $V_{us}$, $V_{ub}$, and $V_{cb}$.
Finally, to match the observed CP phenomenology we included two arbitrary
phases, $\phi_1$, in in $\lambda^{(d)}_{dd}$ and $\phi_2$ in
$\lambda^{(d)}_{ss}$, and computed the Jarlskog invariant
\begin{gather}
J = \Im (V_{us}V_{cb}V^{*}_{ub}V^{*}_{cs})
\end{gather}
and required it to be near the Standard Model expectation $J \approx
3\times 10^{-5}$. The search was performed by Monte Carlo sampling of
the parameter space to find a reasonably close match.

In the lepton sector we are only matching three masses with five
parameters, so solutions are essentially trivial to come by. For
illustration, one such set is
\begin{align}
L_i = 
\begin{pmatrix} 6.118 \\ 5.815 \\ 2.360 \end{pmatrix},\ \ e^+ = 
\begin{pmatrix} 2.696 \\ 3.443 \\ 5.576 \end{pmatrix},
\end{align}
which produces $m_e = 0.511 \mev$ and $m_\mu = 105 \mev$ from matching the
coupling to $m_\tau = 1777 \mev$. There are many others that will match the leptonic masses just as well.

In the quark sector there are many more constraints. It is possible that
there are many configurations that match the Standard Model, but since
this search was for illustrative purposes only the search was stopped
after a solution was obtained. 
This configuration is
\begin{align}
Q_i = 
\begin{pmatrix} 8.132 \\ 2.365 \\ 6.235 \end{pmatrix},\ \ 
\bar{u} = \begin{pmatrix} 9.865 \\ 9.279 \\ 9.404 \end{pmatrix},\ \ 
\bar{d} = \begin{pmatrix} 7.218 \\ 6.463 \\ 8.073 \end{pmatrix},\ \ \ \ 
\phi_1 = -0.0140, \ \ \ \ \phi_2 = -0.0633.
\end{align}
These produce the values
\begin{align}
m_t = 175 \gev &\hspace{2cm} m_b = 4.30 \gev\notag\\
m_c = 1.31 \gev & \hspace{2cm}m_s = 107 \mev\notag\\
m_u = 2.00 \mev & \hspace{2cm}m_d = 8.02 \mev. 
\end{align}
The absolute values of the resulting CKM matrix are
\begin{gather}
|V_{\rm CKM}| = 
\begin{pmatrix}
0.975 & 0.224 & 0.00482\\
0.224 & 0.974 & 0.0439\\
0.00590 & 0.0438 & 0.999
\end{pmatrix}.
\end{gather}
The Jarslkog invariant is $J = 2\times 10^{-5}$. Another way of
estimating the magnitude of CP violation is to calculate
$\sin 2\beta$. Doing this we find $\sin 2\beta = 0.627$, which is less than
$2\sigma$ away from the current world average $\sin 2\beta = 0.734
\pm 0.055$.\cite{Raven:2003gs} Interestingly, under the substitution $\phi_{1,2} \to
-\phi_{1,2}$ all masses and absolute values of CKM elements remain the
same, and $J$ and $\sin 2\beta $ switch sign, picking up the alternate
solution that results from the sign ambiguity in $\beta$. Both of
$J$ and $\sin(2\beta)$ are slightly smaller then the experimental values. However, since this is only an
illustrative configuration, and there may be others that work as well, the
significance is not in exact agreement, but rather in the fact that they
are very close. We note for completeness that this configuration predicts
$\alpha = 2.37$ and $\gamma = 0.430$.

There are a few points to note about this configuration. First, the
largest ratio of widths is $9.865/2.365 \approx 4$, so the widths of all
the fermion fields are of the same order of magnitude, with ${\cal O}(1)$
differences between them. Second, the phases required are very small. That
small phases are required can be easily understood by noting that the
mechanism for generating the hierarchy depends on the singularity of the
Yukawa matrices. A large phase would ruin that singularity. It is thus 
significant that these very small phases are enough to generate the observed
CP-violation.

\section{Flavor Changing Processes}
\label{section:fcnc}

The above discussion of the generation of the Yukawa hierarchy made no
mention of the overall scale of the fermion widths.
This is can be understood from the fact that the Yukawa couplings in
Eq. (\ref{eq:yukawacouplings}) involve only ratios of widths, and hence are
independent of any overall scale.

Effects that depend on the scale will come from interactions with the
Kaluza-Klein (KK) excitations of the gauge bosons. These will turn out to depend explicitly on $R$,
as well as on the ratio $\rho = \sigma/R$. Here $\sigma$ is taken to be
the generic scale of the fermion widths, with $\sigma_i = \gamma_i\sigma$
where $\gamma_i$ is of order unity, and we further define $\rho_i =
\sigma_i/R$ for later convenience. We see that $\rho$ is a measure of the
separation between the energy scales of the fermions (the localization energy)
and the bosons (the compactification scale).

For simplicity at this point we take the extra dimension to be
flat. The bosons are allowed to propagate in the entire bulk, which has
size $R$. The wavefunctions are then 
\begin{gather}
A^{(n)}(y) = \frac{1}{\sqrt{R}}e^{\frac{in\pi y}{R}}\label{eq:bosonwf}.
\end{gather}
In most models half of these will be projected out by boundary
conditions. We retain only the cosine modes, to
preserve the zero modes (n=0) which correspond to the Standard Model
bosons.
The gauge couplings are then
\begin{align}
g^{(n)}_i & = g_5 \int_0^{R}  dy\ A^{(n)}(y) \bar\psi_i(y) \psi_i(y)\notag\\
& = \sqrt{2} g_4 \frac{\left[1-e^{-\frac{1}{\rho_i}}\right]}{1 + \frac{n^2
\pi^2 \rho_i^2}{4}},\hspace{1cm} n \ge 1,\label{eq:gaugecoupling1}
\end{align}
for $a=1$, and 
\begin{gather}
g^{(n)}_i = \sqrt{2} g_{4}e^{-\frac{1}{8}n^2 \rho_i^2}, \hspace{1cm} n \ge
1,\label{eq:gaugecoupling2}
\end{gather}
for $a=2$.

These couplings depend on the fermion species. To see that this results in
flavor-changing currents, consider the
phenomenological 4D Lagrangian for the quarks,
${\boldsymbol{d}}_{L} = (d_{L}, s_{L}, b_{L})^{\dagger}$ and similarly
for ${\bf d}_{R}$, ${\bf u}_{L}$, ${\bf u}_{R}$, all coupled to the gluon
field $G$:
\begin{align}
{\cal L} =&  {\cal L}_{\rm kinetic} + g G^{0}_{\mu}\sum_{i=L,R}
\left(\bar{\boldsymbol{d}}_{i}
\gamma^{\mu} {\boldsymbol{d}}_{i} +  \bar{\boldsymbol{u}}_{i}
\gamma^{\mu} {\boldsymbol{u}}_{i} \right)
 + 
\sqrt{2} g\sum_{n=1}^{\infty} G^{(n)}_{\mu} \sum_{i=L,R}\left(
\bar{\boldsymbol{d}}_i C_{i}^{d(n)} \gamma^{\mu}
{\boldsymbol{d}}_i + \bar{\boldsymbol{u}}_{i} C_{i}^{u(n)}
\gamma^{\mu} {\boldsymbol{u}}_{i} \right)
\notag\\ &
 + 
{\boldsymbol{d}}_{L}V_{L}^{d\dagger}M_d V^d_{R}
{\boldsymbol{d}}_{R}.
+ {\boldsymbol{u}}_{L}V_{L}^{u\dagger}M_u V^u_{R}
{\boldsymbol{u}}_{R},
\end{align}
where $M_{u,d}$ are the diagonal mass matrices, the $V^{(u,d)}_{L,R}$ are
the unitary matrices that accomplish the diagonalization,
the $G^{(n)}$ are the Kaluza-Klein excitations of the gluons,
and we have ignored all but the zero modes for the fermions, since the
fermion KK modes
will not enter the processes considered here. The couplings to the KK
gluons are contained in the diagonal matrices $C_{L,R}^{u,d(n)}$, with
the $i$-th coupling given by Eqs. (\ref{eq:gaugecoupling1}) and
(\ref{eq:gaugecoupling2}). When we transform to the mass basis the gauge
couplings will be the elements of the matrices
\begin{gather}
U^{u,d(n)}_{L,R} \equiv V^{u,d\dagger}_{L,R}C^{u,d(n)}_{L,R}V^{u,d}_{L,R}.
\end{gather}
Since the $C$
matrices are not the identity, these couplings are not diagonal in flavor
space, and hence there will be
flavor-changing currents in the KK gauge sector. In particular there will
be tree-level FCNC mediated by the KK gluon states. 

Processes that involve these tree-level FCNC are the source
of the strong constraints on split fermion models. However, if one
compares the FCNC effects
in our model to the split fermion models, for the same value of the $\rho$
parameter, the magnitude turns out to be much smaller. We can measure the
magnitude of the FCNC with the difference of couplings between two fermion
species $g^{(n)}_i -
g^{(n)}_j$, since the FCNC will vanish if this difference does.
Figure \ref{fig:fcnc} shows the difference of couplings as a function of
the KK number $n$, of two fermion species; in the first case two fermions
of the same width and $\rho =1/10$ separated by $4\sigma$; in the second
two fermions at the same location, one with $\rho_i = 1/10$, the other with
$\rho_j =1/20$. The total effect is the sum over $n$, or roughly the
integral of the curves. Clearly the effect is much smaller in the variable
width case. 

\FIGURE[t]{
\centerline{
\includegraphics[width=6cm,angle=-90]{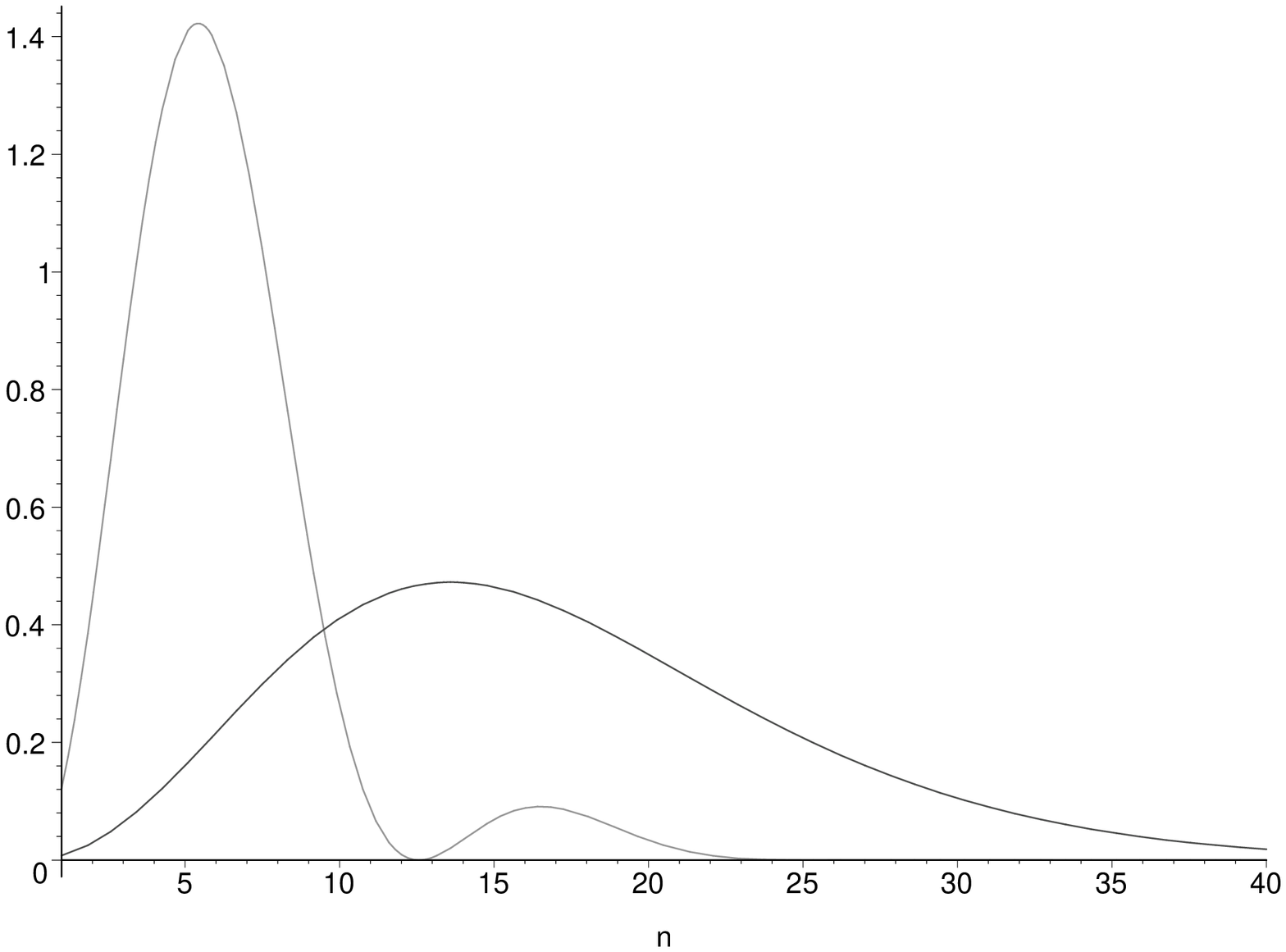}
\includegraphics[width=6cm,angle=-90]{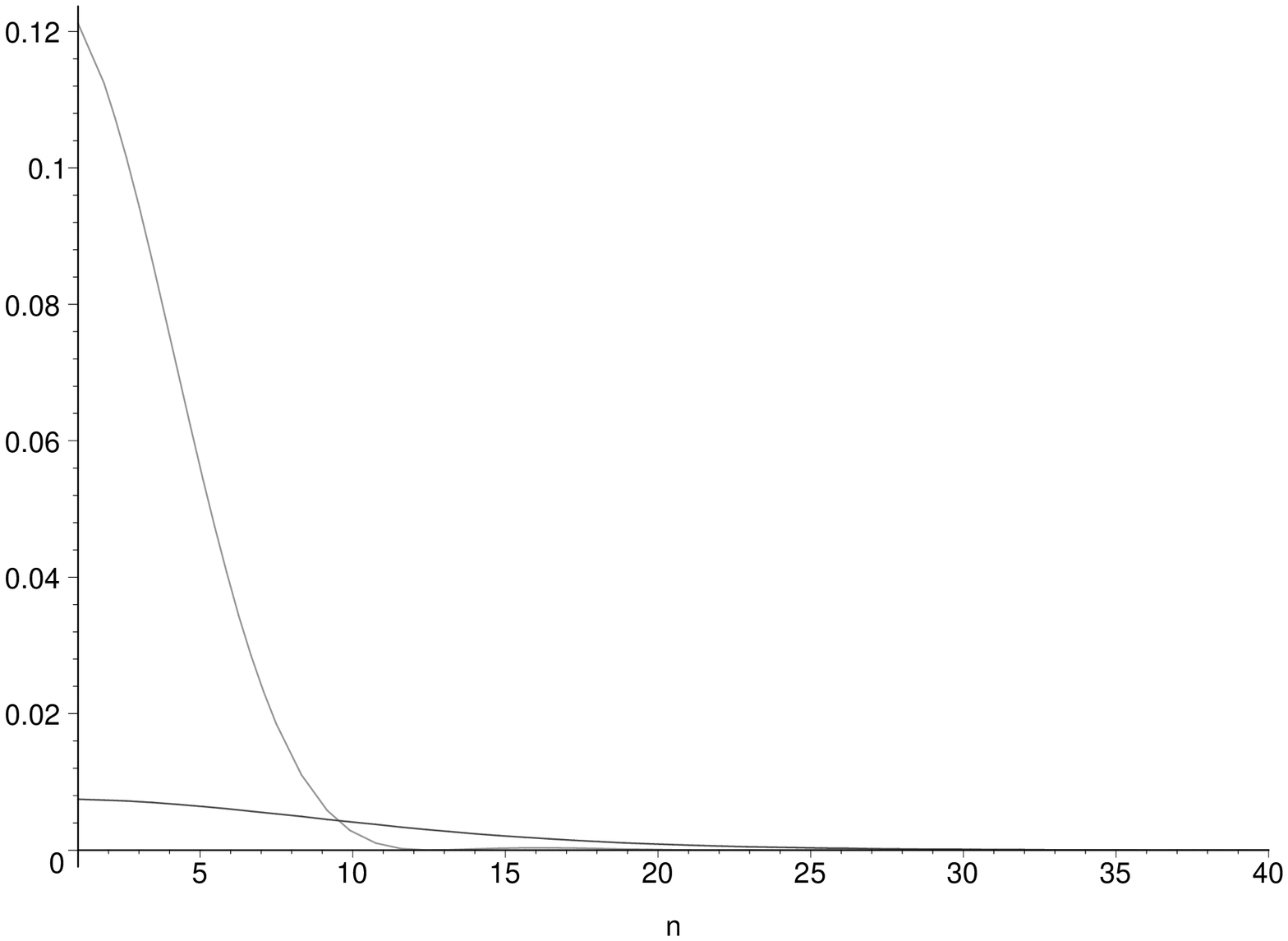}
}
\caption{\small
Comparison of the flavor-changing effects of models with split fermions
(upper curve) and different widths with $a=2$ (lower), for the same
hierarchy of scales, $\sigma/R = 1/10$. Left: the difference of couplings
as a function 
of $n$. Right: the same difference multiplied by $1/n^2$, as it would
appear in the KK sum. Flavor changing currents are proportional to the sum
over $n$ of the right curves.
}
\label{fig:fcnc}}

The reason for this suppression can be understood quite simply. When the fermion
species are directly on top of one another and with the same width the
flavor changing currents are zero. When they are separated the
non-universality of the couplings comes from the different heights of the
KK wavefunctions at the location of the fermions. So every KK state,
starting from $n=1$, can resolve the difference, up to the cutoff $n \approx
1/\rho$. When the fermions are localized to the same point and the widths
are changed the
non-universality can only start to be resolved when the KK wavefunctions
oscillate fast enough to resolve the widths. In this case the first few KK
states will have nearly universal couplings, and the large flavor
difference won't start until roughly $n \approx 1/\rho_>$, where
$\rho_>$ is the larger of the two widths, and will only last until
$n\approx 1/\rho_<$. Additionally, if there is only one extra dimension,
the mass in the KK propagator provides a $1/n^2$ suppression in the sum,
further
suppressing the FCNC from variable widths, while leaving the FCNC from the
first few KK modes from split fermions large. 
Note that this argument does not depend crucially on the shape of the
fermion wavefunctions. The key points are the slow variation of the gauge
wavefunctions for small $n$ over the size of the fermions, and the $1/n^2$
suppression in the propagator. Thus the argument holds for any
value of $a$, and indeed would hold for very different, potentially very
irregular, shapes of the fermions. The only requirement is that they all fall
off exponentially or faster from the same point on length scales of order
the first KK mass.

The most stringent specific constraint comes from the mass splitting in
the neutral kaon sector, $\Delta m_K$. The calculation of $\Delta m_K$
here is identical with that in the split fermion case, except for the form
of the sum over KK-modes. In \cite{Lillie:2003yz} it was found that
\begin{align}
\Delta m_K = 
&\Re\langle \bar K^0|{\cal L}^{\Delta S =2}|K^0\rangle
\notag\\ 
= \frac{2}{3}R^2 f_K^2 m_K
&\Re\left(\vphantom{\frac{1}{1}}\right.
|V^{d}_{L\ 11}V^{d*}_{L\ 12}|^2 \zeta_1
\sum_{n=1}^{\infty}\frac{(g_{d_L} - g_{s_L})^2}{n^2}
+|V^{d}_{R\ 11}V^{d*}_{R\ 12}|^2\zeta_1
\sum_{n=1}^{\infty}\frac{(g_{d_R} - g_{s_R})^2}{n^2}
\label{eq:ds2}\\
&+(V^{d}_{L\ 11}V^{d*}_{L\ 12}V^{d*}_{R\ 11}V^{d}_{R\ 12})\zeta_2
\sum_{n=1}^{\infty}\frac{(g_{d_L} - g_{s_L})(g_{d_R} - g_{s_R})}{n^2}
\notag\\
&+(V^{d}_{R\ 11}V^{*d}_{R\ 12}V^{*d}_{L\ 11}V^{d}_{L\ 12})\zeta_2
\sum_{n=1}^{\infty}\frac{(g_{d_L} - g_{s_L})(g_{d_R} - g_{s_R})}{n^2}
\left.\vphantom{\frac{1}{1}}\right),\notag
\end{align}
where the $\zeta_i$ are the dimensionless part of the hadronic matrix
elements and are given by
\begin{align}
\zeta_1 & = \frac{1}{3}, \notag\\
\zeta_2 & = \left(
\frac{1}{12} + \frac{1}{4}\left(\frac{m_K^2}{m_d^2 + m_s^2}
\right)\right),
\end{align}
when computed in the vacuum insertion approximation \cite{Gabbiani:1996hi}.
Requiring this to not be larger than the observed
value produces a constraint on $1/R$, given a value of
$\rho$, 
\begin{gather}
\frac{1}{R} \ge 1960 \tev \sqrt{\Re\left(\zeta_1 \sum({\rm LL} + {\rm RR}) + \zeta_2
\sum({\rm LR} + {\rm RL})\right)}\label{eq:deltamc},
\end{gather}
where ${\rm LL}$ is the left-left term in Eq. (\ref{eq:ds2}), etc..

We have calculated $\Delta m_K$ for the configuration given in section
\ref{section:standardmodel}.
Fig. \ref{fig:deltamk} shows the constraint for a range of
$\rho$. This demonstrates clear power-law behavior, with $1/R = (73.8\tev)
\rho^{1/2}$. We see that for very large values, say $\rho =10^{-1}$, we
have large constraints $1/R \ge 30 \tev$. One can get down to the direct
production constraint of $1/R \ge 2 \tev$ by going to $\rho \approx
10^{-3}$.
The constraints here turn out to be very close to
those obtained for the model of Kaplan and Tait \cite{Kaplan:2001ga},
where they localized fermions to exponentials centered at one of two fixed
points, much like here, but still produced small Yukawa matrix elements by
separating the left and right handed components.\footnote{The paper
\cite{Kaplan:2001ga} presents results for $\rho = 10^{-1}$. The apparent
disagreement between the numbers quoted there and here is due to a
normalization factor of $2\pi$ between their $M_c$ and our $1/R$.} Since
they presumably have very different $V_{L,R}^{u,d}$ mixing elements, this
shows that the flavor constraints on variable width models are quite robust.
These constraints are to be contrasted with the similar results for the split
fermion case, where $1/R \ge 400 \tev$ for $\rho = 10^{-1}$ and $1/R \ge
60 \tev$ for $\rho = 10^{-3}$; in those models one must go to $\rho
\approx 10^{-5}$ before the flavor constraints are similar to direct
production constraints.

\FIGURE[t]{
\centerline{
\includegraphics[width=12cm]{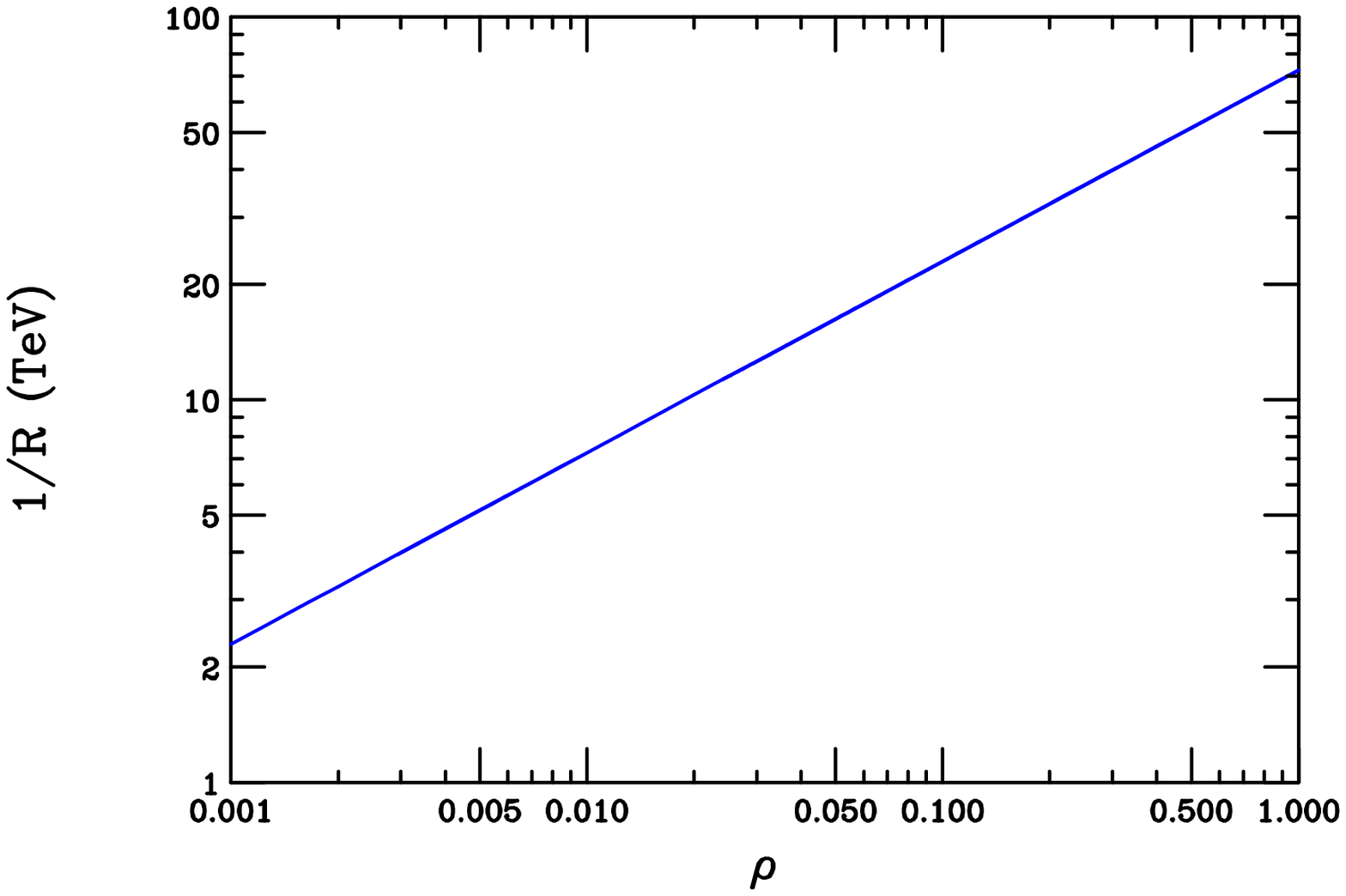}
}
\caption{\small
Constraints on the compactification scale $1/R$ from $\Delta m_K$ as a
function of $\rho = \sigma/R$, the ratio of the compactification scale to
the fermion localization scale. The area below the curve is excluded.
}
\label{fig:deltamk}}

Note that in several places it was claimed that the constraint from
$\epsilon_K$ was even larger than that from $\Delta m_K$ \cite{Delgado:1999sv,Abel:2003fk,Kaplan:2001ga}. This can
be seen from the $\epsilon_K$ equivalent to Eq. (\ref{eq:deltamc})
\begin{gather}
\frac{1}{R} \ge 40,900 \tev \sqrt{\Im\left(\zeta_1 \sum({\rm LL} + {\rm RR}) + \zeta_2
\sum({\rm LR} + {\rm RL})\right)}\label{eq:epsilonc}
\end{gather}
However, the imaginary part is, in fact, quite small. For $\rho =
10^{-1}$, the factor in the square root is $1.6\times 10^{-3}$, leading to
a constraint $1/R \ge 6.6 \tev$; this is much smaller than the constraint from
$\Delta m_K$.

\section{Discussion}
\label{section:mechanisms}

In previous sections we have tried to emphasize the aspects which are
independent of the mechanism for the variable width scenario of fermion
localization. 
Here we discuss possible techniques of fermion localization that produce
this scenario. The easiest implementation of localized fermions is on a
five-dimensional space with the extra dimension compactified on $S_1/Z_2$.
The fermions are then coupled to a scalar field that is odd under the
$Z_2$ action of the orbifold. It can then be shown that the fermions
will develop chiral zero modes localized near one of the orbifold fixed
points \cite{Georgi:2000wb}. This Lagrangian is given by
\begin{gather}
{\cal L} = \sum_i \bar \Psi_i(i\not\! \partial - \gamma^5\partial_5 - f_i
\varphi)\Psi_i +\frac{1}{2}\partial^{\mu}\varphi \partial_\mu \varphi
- \frac{1}{2}\partial_5 \varphi \partial_5 \varphi
-\frac{\lambda}{4}(\varphi^2 - v^2)^2,
\end{gather}
where here we allow each fermion to have a separate coupling, $f_i$, to
$\varphi$. Since $\varphi$ is odd under the orbifolding it must vanish at
each of the fixed points, $y=0$ and $y=\pi R$. However, if $\lambda v^2$
is large enough
$\varphi$ will develop a $y$-dependent vev, $h(y)$, that is zero on the
fixed points and is non-vanishing elsewhere. The
fermions then develop zero modes with profiles
\begin{gather}
\psi_i(y) \propto e^{-f_i \int_0^{y} dy'\ h(y')}.
\end{gather}
This is localized near $y=0$ if $f_i h(y) > 0$, and at $y=R$
otherwise. The class of models parameterized by $a$ in
Eq. (\ref{eq:profile}) can be constructed by demanding that $h(y)$
behave like $h(y) \approx k y^{a-1}$ near $y=0$ for some constant $k$, and
picking all the $f$ to have the same sign as $h(y)$, so all fermions are
localized to $y=0$. The fermions are then localized as in
Eq. (\ref{eq:profile}) with width parameter $\sigma_i = (f_i k)^{-1/a}$.

Note that this construction does not appear to generate any non-trivial
phases in the Yukawa matrices with which we could generate
$\phi_{1,2}$. However, phases will be present if one allows the localizer
field $\varphi$ to be complex and imposing boundary conditions on the
phase at the orbifold fixed points. For instance we could require that
$\arg(\varphi(y=0)) = 0$ and $\arg(\varphi(y=R)) = \pi/2$. The vev
then has a profile $ve^{i\frac{\pi}{2}\frac{y}{R}}$. This still results
in localized fermions, but the Yukawa matrix elements pick up phases
relative to each other, since the phase at each point will be weighted
differently due to the distinct widths. For fermions with width ratios of
order those given in the solution in section \ref{section:standardmodel} with
$1/R \approx 5 \tev$, we find phase differences of order $\phi = 0.03$, so
this looks like a promising way to make a realistic construction of this
scenario.

Another interesting example of fermion localization occurs in the
RS scenario. In this case, the
fermions are localized near a brane with $a=1$ profiles, and possibly
different widths \cite{Hewett:2002fe}. The gauge boson wavefunctions are
Bessel functions rather than cosines, so the
flavor analysis in section \ref{section:fcnc} is not strictly
applicable. However, since the low KK-number wavefunctions are reasonably
flat, the reasoning that the flavor changing currents are small
will still hold. This has been seen explicitly in \cite{Huber:2003tu}.

One would like to embed the variable widths model in a larger scenario
that, for instance, solves the Higgs mass hierarchy problem. The fermions
and gauge bosons 
could be localized within a thick brane embedded in a larger dimension of
the ADD type. Or the complete manifold could be $AdS_5 \times (S_1/Z_2)$ with
the RS scenario playing out in the $AdS_5$ and the fermion mass generation playing
out in the $S_1/Z_2$ \cite{Davoudiasl:2002wz}. Unfortunately, with all fermions
localized at a
single point there is nothing to suppress the proton decay operators that
tend to occur in these models, unlike in the original split fermion
scenario. However, a simple twist on this scenario can restore this suppression. Instead of a single
fixed point, there could be two points, with the quark fields localized on
one of them, and the leptons on the other, as shown in
Fig. \ref{fig:quarksandleptons}. Proton decay operators are then
suppressed by approximately $e^{-\frac{3}{4} l^2}$, where $l$ is the
separation between the two
fixed points, in units of the scale of the fermion widths. Note
that this changes none of the conclusions above about the generation of the
mass hierarchy. In the simple $S_1/Z_2$ model presented above this could
be accomplished by taking the couplings to $\varphi$ for quarks to be $f_i > 0$ and for
leptons $f_i < 0$.

\FIGURE[t]{
\centerline{
\includegraphics[width=6cm,height=10cm,angle=-90]{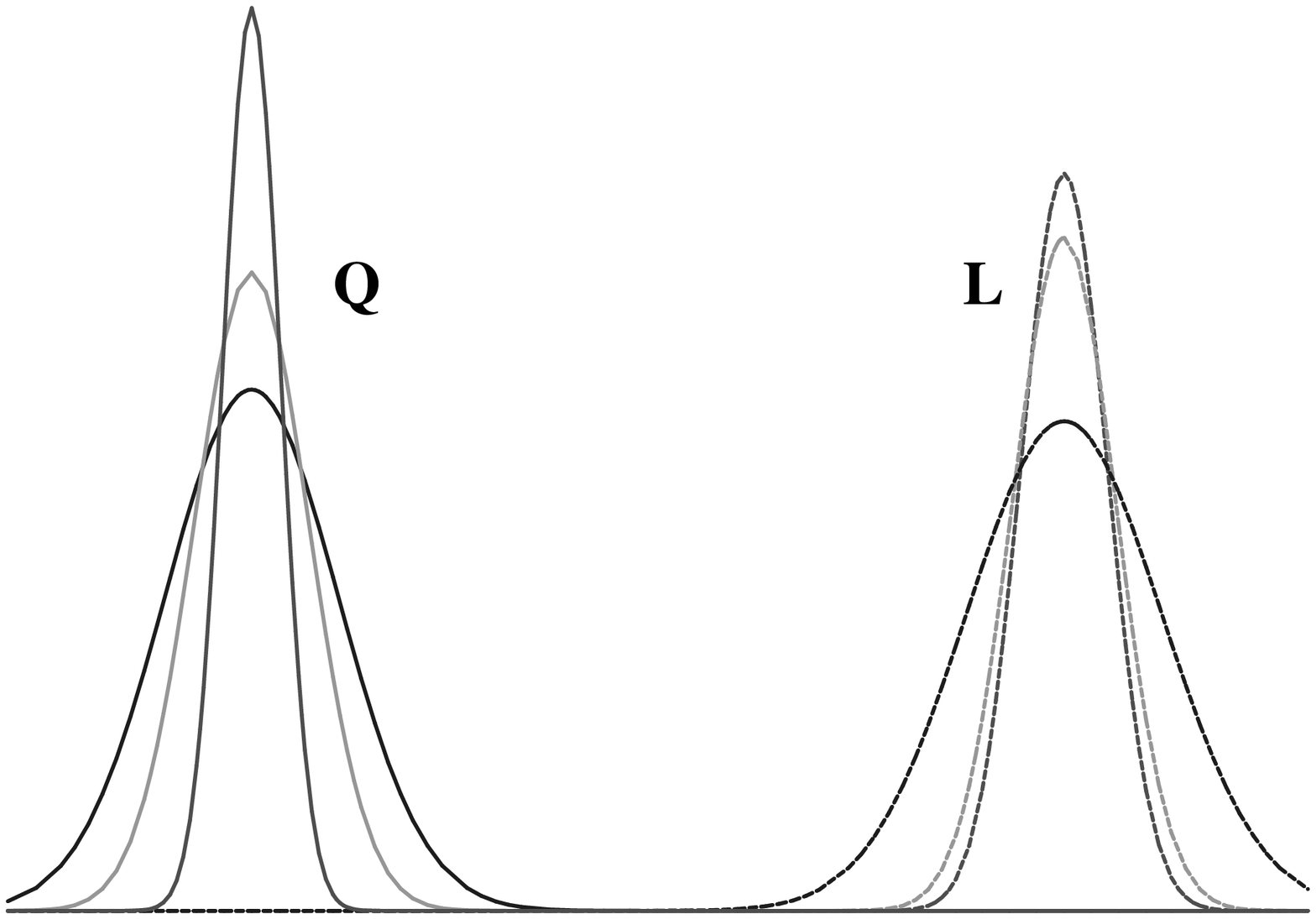}}
\caption{\small
Illustration of a configuration that would suppress proton decay
while generating the Yukawa hierarchy with variable widths.
}
\label{fig:quarksandleptons}}

We could also try to modify the model in order to generate the ratios
$\lambda_5^u/\lambda_5^d$ and $\lambda_5^u/\lambda_5^e$, rather than
having to put them in by hand. One way to do this would be to make use of
the original AS suggestion of separating left and right handed
fermions. The up-type singlets, $u_i$, could be very close to the quark
doublets, $Q_i$. The down-type singlets, $d_i$, could then be farther away
(about $2\sigma$ for $a=2$), and the lepton doublets and charged singlets
separated slightly farther than that. This is then a hybrid model where the
hierarchies between fermions with the same charge are generated by the
different widths, and the hierarchy between those with different charges is generated
by the exponentially small overlaps. Combined with separating the lepton
wavefunctions to suppress
proton decay, one obtains the picture in Fig. \ref{fig:everything}. The whole assembly
with $a=2$ requires about $10\sigma$ of length in the additional
dimension; most of that is needed to suppress proton
decay.
Note that as long as all fermions with the same quantum numbers are
localized to the same place the suppression of FCNC described above will
apply.

\FIGURE[t]{
\centerline{
\includegraphics[width=14cm,height=7cm]{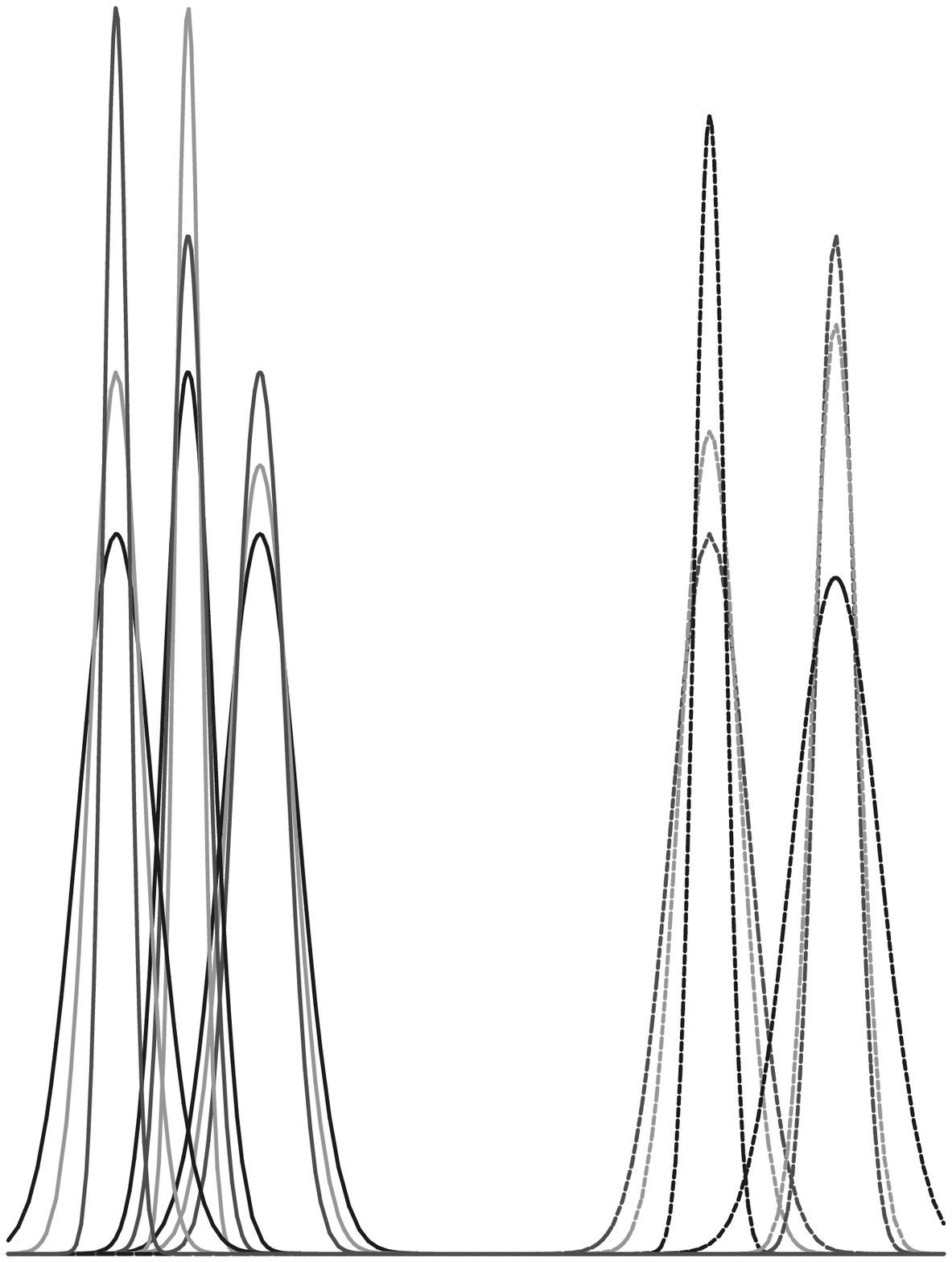}}
\caption{\small
Illustration showing a ``Swiss Army Knife'' configuration. It generates the
hierarchy between generations with the variable width method; uses split
fermions to generate the hierarchy between the top mass and the bottom and
$\tau$ masses; and suppresses proton decay by localizing quarks and
fermions to different fixed points.
}
\label{fig:everything}}

\section{Conclusion}
\label{section:conclusion}

We have proposed a model that localizes fermions with different widths in
a single, compact, extra dimensional bulk, and shown that this
produces a realization of the democratic scenario of fermion mass
matrices. Ratios of the fermions widths that
are ${\cal O}(1)$ can produce the observed fermion masses and mixing
angles. With the inclusion of additional small
phases one can reproduce all Standard Model parameters in the fermion
sector, including the CP-violating CKM phase. This is not a reduction of
parameters, but it does explain the large hierarchy in Yukawa
couplings in terms of a simple physical picture. The tree-level FCNC
contributions are smaller than in similar models
by about a factor of $10$ per flavor changing vertex. The resulting
constraints on the compactification scale can be as small as $1-2 \tev$ if
one allows the ratio of localization to compactification scales
to be $\rho \approx 10^{-3}$.
This advantage in reducing FCNC effects comes not so much from the
particular mechanism of localization, but rather from the facts that all
fermions are localized to the same point, and that the gauge boson
KK wavefunctions are very smooth for the first few modes, so they fail to
resolve the difference in fermion zero modes. When the KK mode number $n$
is high
enough to resolve the differences in fermions the $1/n^2$ suppression is
large. It is also interesting that the mechanism
for generating the Yukawa hierarchy does not depend on the shape of the
fermion wavefunction near the localization point, as long as it is
exponential far away. It only requires ${\cal O}(1)$ differences between
overlaps of the wavefunctions. As long as they were all appropriately
localized, the zero-mode wavefunctions could be extremely irregular and
still generate the correct fermion hierarchy while being consistent with
the FCNC constraints.
It may also be possible to reduce the FCNC constraints further by the
inclusion of the effects of brane-localized kinetic terms, which were not
considered here \cite{Carena:2002me,delAguila:2003bh}.

Effects of variable width localization could be observable at future
colliders, particularly in flavor-changing
processes. Even if direct effects are not seen at the LHC, precision
measurements of mass splittings or rare $B$ decays may provide clues. In
\cite{Arkani-Hamed:1999za}, it was noted
that when fermions are separated in an extra dimension, one might be able
to observe cross sections that fall exponentially in the center-of-mass
energy in certain channels. There it was proposed to look at polarized
$e^+e^-$ collisions to search for split fermions. In the scenario
presented here this may apply to high energy $ep$ collisions if the quark
and lepton separation prevents proton decay.

\bigskip

\noindent
{\bf Note Added:} While this paper was being completed a similar paper
\cite{Soddu:2003zz} appeared. There, a specific model for the vev in section
\ref{section:mechanisms} was made and the Standard
Model parameters were obtained.

\acknowledgments

The author would like to thank Tom Rizzo, Frank Petriello, JoAnne
Hewett, and Michael Peskin for many useful discussions.

\bibliography{semisplit}

\providecommand{\href}[2]{#2}\begingroup\raggedright\begin{thebibliography}{10}

\bibitem{Antoniadis:1990ew}
I.~Antoniadis, {\it A possible new dimension at a few tev},  {\em Phys. Lett.}
  {\bf B246} (1990) 377--384.

\bibitem{Lykken:1996fj}
J.~D. Lykken, {\it Weak scale superstrings},  {\em Phys. Rev.} {\bf D54} (1996)
  3693--3697, [\href{http://xxx.lanl.gov/abs/hep-th/9603133}{{\tt
  hep-th/9603133}}].

\bibitem{Witten:1996mz}
E.~Witten, {\it Strong coupling expansion of calabi-yau compactification},
  {\em Nucl. Phys.} {\bf B471} (1996) 135--158,
  [\href{http://xxx.lanl.gov/abs/hep-th/9602070}{{\tt hep-th/9602070}}].

\bibitem{Horava:1996ma}
P.~Horava and E.~Witten, {\it Eleven-dimensional supergravity on a manifold
  with boundary},  {\em Nucl. Phys.} {\bf B475} (1996) 94--114,
  [\href{http://xxx.lanl.gov/abs/hep-th/9603142}{{\tt hep-th/9603142}}].

\bibitem{Horava:1996qa}
P.~Horava and E.~Witten, {\it Heterotic and type i string dynamics from eleven
  dimensions},  {\em Nucl. Phys.} {\bf B460} (1996) 506--524,
  [\href{http://xxx.lanl.gov/abs/hep-th/9510209}{{\tt hep-th/9510209}}].

\bibitem{Caceres:1997is}
E.~Caceres, V.~S. Kaplunovsky, and I.~M. Mandelberg, {\it Large-volume string
  compactifications, revisited},  {\em Nucl. Phys.} {\bf B493} (1997) 73--100,
  [\href{http://xxx.lanl.gov/abs/hep-th/9606036}{{\tt hep-th/9606036}}].

\bibitem{Arkani-Hamed:1998rs}
N.~Arkani-Hamed, S.~Dimopoulos, and G.~R. Dvali, {\it The hierarchy problem and
  new dimensions at a millimeter},  {\em Phys. Lett.} {\bf B429} (1998)
  263--272, [\href{http://xxx.lanl.gov/abs/hep-ph/9803315}{{\tt
  hep-ph/9803315}}].

\bibitem{Antoniadis:1998ig}
I.~Antoniadis, N.~Arkani-Hamed, S.~Dimopoulos, and G.~R. Dvali, {\it New
  dimensions at a millimeter to a fermi and superstrings at a tev},  {\em Phys.
  Lett.} {\bf B436} (1998) 257--263,
  [\href{http://xxx.lanl.gov/abs/hep-ph/9804398}{{\tt hep-ph/9804398}}].

\bibitem{Arkani-Hamed:1998nn}
N.~Arkani-Hamed, S.~Dimopoulos, and G.~R. Dvali, {\it Phenomenology,
  astrophysics and cosmology of theories with sub-millimeter dimensions and tev
  scale quantum gravity},  {\em Phys. Rev.} {\bf D59} (1999) 086004,
  [\href{http://xxx.lanl.gov/abs/hep-ph/9807344}{{\tt hep-ph/9807344}}].

\bibitem{Randall:1999ee}
L.~Randall and R.~Sundrum, {\it A large mass hierarchy from a small extra
  dimension},  {\em Phys. Rev. Lett.} {\bf 83} (1999) 3370--3373,
  [\href{http://xxx.lanl.gov/abs/hep-ph/9905221}{{\tt hep-ph/9905221}}].

\bibitem{Randall:1999vf}
L.~Randall and R.~Sundrum, {\it An alternative to compactification},  {\em
  Phys. Rev. Lett.} {\bf 83} (1999) 4690--4693,
  [\href{http://xxx.lanl.gov/abs/hep-th/9906064}{{\tt hep-th/9906064}}].

\bibitem{Dobrescu:2001ae}
B.~A. Dobrescu and E.~Poppitz, {\it Number of fermion generations derived from
  anomaly cancellation},  {\em Phys. Rev. Lett.} {\bf 87} (2001) 031801,
  [\href{http://xxx.lanl.gov/abs/hep-ph/0102010}{{\tt hep-ph/0102010}}].

\bibitem{Appelquist:2000nn}
T.~Appelquist, H.-C. Cheng, and B.~A. Dobrescu, {\it Bounds on universal extra
  dimensions},  {\em Phys. Rev.} {\bf D64} (2001) 035002,
  [\href{http://xxx.lanl.gov/abs/hep-ph/0012100}{{\tt hep-ph/0012100}}].

\bibitem{Servant:2002aq}
G.~Servant and T.~M.~P. Tait, {\it Is the lightest kaluza-klein particle a
  viable dark matter candidate?},  {\em Nucl. Phys.} {\bf B650} (2003)
  391--419, [\href{http://xxx.lanl.gov/abs/hep-ph/0206071}{{\tt
  hep-ph/0206071}}].

\bibitem{Servant:2002hb}
G.~Servant and T.~M.~P. Tait, {\it Elastic scattering and direct detection of
  kaluza-klein dark matter},  {\em New J. Phys.} {\bf 4} (2002) 99,
  [\href{http://xxx.lanl.gov/abs/hep-ph/0209262}{{\tt hep-ph/0209262}}].

\bibitem{Cheng:2002ej}
H.-C. Cheng, J.~L. Feng, and K.~T. Matchev, {\it Kaluza-klein dark matter},
  {\em Phys. Rev. Lett.} {\bf 89} (2002) 211301,
  [\href{http://xxx.lanl.gov/abs/hep-ph/0207125}{{\tt hep-ph/0207125}}].

\bibitem{Arkani-Hamed:1999dc}
N.~Arkani-Hamed and M.~Schmaltz, {\it Hierarchies without symmetries from extra
  dimensions},  {\em Phys. Rev.} {\bf D61} (2000) 033005,
  [\href{http://xxx.lanl.gov/abs/hep-ph/9903417}{{\tt hep-ph/9903417}}].

\bibitem{Grossman:2002pb}
Y.~Grossman and G.~Perez, {\it Realistic construction of split fermion models},
   {\em Phys. Rev.} {\bf D67} (2003) 015011,
  [\href{http://xxx.lanl.gov/abs/hep-ph/0210053}{{\tt hep-ph/0210053}}].

\bibitem{Mirabelli:1999ks}
E.~A. Mirabelli and M.~Schmaltz, {\it Yukawa hierarchies from split fermions in
  extra dimensions},  {\em Phys. Rev.} {\bf D61} (2000) 113011,
  [\href{http://xxx.lanl.gov/abs/hep-ph/9912265}{{\tt hep-ph/9912265}}].

\bibitem{Chang:2002ww}
W.-F. Chang and J.~N. Ng, {\it Cp violation in 5d split fermions scenario},
  {\em JHEP} {\bf 12} (2002) 077,
  [\href{http://xxx.lanl.gov/abs/hep-ph/0210414}{{\tt hep-ph/0210414}}].

\bibitem{DelAguila:2001pu}
F.~Del~Aguila and J.~Santiago, {\it Signals from extra dimensions decoupled
  from the compactification scale},  {\em JHEP} {\bf 03} (2002) 010,
  [\href{http://xxx.lanl.gov/abs/hep-ph/0111047}{{\tt hep-ph/0111047}}].

\bibitem{Kaplan:2001ga}
D.~E. Kaplan and T.~M.~P. Tait, {\it New tools for fermion masses from extra
  dimensions},  {\em JHEP} {\bf 11} (2001) 051,
  [\href{http://xxx.lanl.gov/abs/hep-ph/0110126}{{\tt hep-ph/0110126}}].

\bibitem{Branco:2000rb}
G.~C. Branco, A.~de~Gouvea, and M.~N. Rebelo, {\it Split fermions in extra
  dimensions and cp violation},  {\em Phys. Lett.} {\bf B506} (2001) 115--122,
  [\href{http://xxx.lanl.gov/abs/hep-ph/0012289}{{\tt hep-ph/0012289}}].

\bibitem{Rizzo:1999br}
T.~G. Rizzo and J.~D. Wells, {\it Electroweak precision measurements and
  collider probes of the standard model with large extra dimensions},  {\em
  Phys. Rev.} {\bf D61} (2000) 016007,
  [\href{http://xxx.lanl.gov/abs/hep-ph/9906234}{{\tt hep-ph/9906234}}].

\bibitem{Masip:1999mk}
M.~Masip and A.~Pomarol, {\it Effects of sm kaluza-klein excitations on
  electroweak observables},  {\em Phys. Rev.} {\bf D60} (1999) 096005,
  [\href{http://xxx.lanl.gov/abs/hep-ph/9902467}{{\tt hep-ph/9902467}}].

\bibitem{Marciano:1999ih}
W.~J. Marciano, {\it Fermi constants and 'new physics'},  {\em Phys. Rev.} {\bf
  D60} (1999) 093006, [\href{http://xxx.lanl.gov/abs/hep-ph/9903451}{{\tt
  hep-ph/9903451}}].

\bibitem{Hewett:2002hv}
J.~Hewett and M.~Spiropulu, {\it Particle physics probes of extra spacetime
  dimensions},  {\em Ann. Rev. Nucl. Part. Sci.} {\bf 52} (2002) 397--424,
  [\href{http://xxx.lanl.gov/abs/hep-ph/0205106}{{\tt hep-ph/0205106}}].

\bibitem{Arkani-Hamed:1999za}
N.~Arkani-Hamed, Y.~Grossman, and M.~Schmaltz, {\it Split fermions in extra
  dimensions and exponentially small cross-sections at future colliders},  {\em
  Phys. Rev.} {\bf D61} (2000) 115004,
  [\href{http://xxx.lanl.gov/abs/hep-ph/9909411}{{\tt hep-ph/9909411}}].

\bibitem{Rizzo:2001cy}
T.~G. Rizzo, {\it Cartography with accelerators: Locating fermions in extra
  dimensions at future lepton colliders},  {\em Phys. Rev.} {\bf D64} (2001)
  015003, [\href{http://xxx.lanl.gov/abs/hep-ph/0101278}{{\tt
  hep-ph/0101278}}].

\bibitem{Delgado:1999sv}
A.~Delgado, A.~Pomarol, and M.~Quiros, {\it Electroweak and flavor physics in
  extensions of the standard model with large extra dimensions},  {\em JHEP}
  {\bf 01} (2000) 030, [\href{http://xxx.lanl.gov/abs/hep-ph/9911252}{{\tt
  hep-ph/9911252}}].

\bibitem{Abel:2003fk}
S.~A. Abel, M.~Masip, and J.~Santiago, {\it Flavour changing neutral currents
  in intersecting brane models},
  \href{http://xxx.lanl.gov/abs/hep-ph/0303087}{{\tt hep-ph/0303087}}.

\bibitem{Lillie:2003yz}
B.~Lillie and J.~Hewett, {\it Flavor constraints on split fermion models},
  \href{http://xxx.lanl.gov/abs/hep-ph/0306193}{{\tt hep-ph/0306193}}.

\bibitem{Huber:2003tu}
S.~J. Huber, {\it Flavor violation and warped geometry},
  \href{http://xxx.lanl.gov/abs/hep-ph/0303183}{{\tt hep-ph/0303183}}.

\bibitem{Fritzsch:1999ee}
H.~Fritzsch and Z.-z. Xing, {\it Mass and flavor mixing schemes of quarks and
  leptons},  {\em Prog. Part. Nucl. Phys.} {\bf 45} (2000) 1--81,
  [\href{http://xxx.lanl.gov/abs/hep-ph/9912358}{{\tt hep-ph/9912358}}].

\bibitem{Hagiwara:2002fs}
{\bf Particle Data Group} Collaboration, K.~Hagiwara {\em et.~al.}, {\it Review
  of particle physics},  {\em Phys. Rev.} {\bf D66} (2002) 010001.

\bibitem{Raven:2003gs}
G.~Raven, {\it sin(2beta): Status and prospects},
  \href{http://xxx.lanl.gov/abs/hep-ex/0307067}{{\tt hep-ex/0307067}}.

\bibitem{Gabbiani:1996hi}
F.~Gabbiani, E.~Gabrielli, A.~Masiero, and L.~Silvestrini, {\it A complete
  analysis of fcnc and cp constraints in general susy extensions of the
  standard model},  {\em Nucl. Phys.} {\bf B477} (1996) 321--352,
  [\href{http://xxx.lanl.gov/abs/hep-ph/9604387}{{\tt hep-ph/9604387}}].

\bibitem{Georgi:2000wb}
H.~Georgi, A.~K. Grant, and G.~Hailu, {\it Chiral fermions, orbifolds, scalars
  and fat branes},  {\em Phys. Rev.} {\bf D63} (2001) 064027,
  [\href{http://xxx.lanl.gov/abs/hep-ph/0007350}{{\tt hep-ph/0007350}}].

\bibitem{Hewett:2002fe}
J.~L. Hewett, F.~J. Petriello, and T.~G. Rizzo, {\it Precision measurements and
  fermion geography in the randall-sundrum model revisited},  {\em JHEP} {\bf
  09} (2002) 030, [\href{http://xxx.lanl.gov/abs/hep-ph/0203091}{{\tt
  hep-ph/0203091}}].

\bibitem{Davoudiasl:2002wz}
H.~Davoudiasl, J.~L. Hewett, and T.~G. Rizzo, {\it Phenomenology on a slice of
  ads(5) x m**delta spacetime},  {\em JHEP} {\bf 04} (2003) 001,
  [\href{http://xxx.lanl.gov/abs/hep-ph/0211377}{{\tt hep-ph/0211377}}].

\bibitem{Carena:2002me}
M.~Carena, T.~M.~P. Tait, and C.~E.~M. Wagner, {\it Branes and orbifolds are
  opaque},  {\em Acta Phys. Polon.} {\bf B33} (2002) 2355,
  [\href{http://xxx.lanl.gov/abs/hep-ph/0207056}{{\tt hep-ph/0207056}}].

\bibitem{delAguila:2003bh}
F.~del Aguila, M.~Perez-Victoria, and J.~Santiago, {\it Bulk fields with
  general brane kinetic terms},  {\em JHEP} {\bf 02} (2003) 051,
  [\href{http://xxx.lanl.gov/abs/hep-th/0302023}{{\tt hep-th/0302023}}].

\bibitem{Soddu:2003zz}
A.~Soddu and N.-K. Tran, {\it Democratic mass matrices from five dimensions},
  \href{http://xxx.lanl.gov/abs/hep-ph/0308043}{{\tt hep-ph/0308043}}.

\end{thebibliography}\endgroup

\end{document}